\documentstyle[12pt]{article}
\input{psbox.tex}
\normalsize

\def\a{\alpha}

\def\d{\delta}

\def\j{\psi}

\def\Hat#1{\rlap{\kern.10em$\widehat{\phantom G}$}#1}
\def\HAt#1{\rlap{\kern.05em$\widehat{\phantom G}$}#1}

\def\cap#1{\rlap{\kern.1em$\widehat{\phantom{G\vrule height.8em}}$}#1{}}
\def\Cap#1{\rlap{\kern.05em$\widehat{\phantom{G\vrule height.8em}}$}#1{}}

\newcounter{sxn}

\newcounter{axn}

\def\br{\begin{thebibliography}{99}}
\def\er{\end{thebibliography}}

\date{}

\begin{document}
\bibliographystyle{unsrt}
\footskip 1.0cm
\thispagestyle{empty}
\setcounter{page}{0}
\begin{flushright}
SU-4228-533\\
March 1993\\
\end{flushright}
\vspace{10mm}
\centerline{\LARGE TOPOLOGY IN PHYSICS - A PERSPECTIVE}

\vspace*{5mm}
\centerline {\large A.P. Balachandran}
\vspace*{5mm}
\centerline {\it Department of Physics, Syracuse University,}
\centerline {\it Syracuse, NY 13244-1130}
\vspace*{25mm}
\normalsize
\centerline {\large ABSTRACT}
\vspace*{5mm}
This article, written in honour of Fritz Rohrlich, briefly surveys the role of
topology in physics.
\newpage

\baselineskip=24pt
\setcounter{page}{1}

When I joined Syracuse University as a junior faculty member in September,
1964,
 Fritz
Rohrlich was already there as a senior theoretician in the quantum field theory
group.  He was a very well-known physicist by that time, having made
fundamental contributions to quantum field theory, and written his splendid
book with Jauch.  My generation of physicists grew up with Jauch and Rohrlich,
and I was also familiar with Fritz's research.  It was therefore with a certain
awe and a great deal of respect that I first made his acquaintance.

Many years have passed since this first encounter.  Our interests too have
gradually evolved and changed in this intervening time.  Starting from the late
seventies or thereabouts, particle physicists have witnessed an increasing
intrusion of topological ideas into their discipline.  Our group at Syracuse
has responded to this development by getting involved in soliton and monopole
physics and in investigations on the role of topology in quantum physics.
Meanwhile, especially during the last decade, there has been a perceptible
shift in the direction of Fritz's research to foundations and history of
quantum physics.  Later I will argue that topology affects the nature of wave
functions (or more accurately of wave functions in the domains of observables)
and has a profound meaning for the fundamentals of quantum theory.  It may not
therefore be out of place to attempt a partly historical and occasionally
technical essay on topology in physics for the purpose of dedication to Fritz.

This article has no pretension to historical accuracy or scholarship.  As
alluded to previously, particle theorists have come to appreciate the
importance of topology in the classical and especially in the quantum domain
over the years, and that has evoked a certain curiosity about its role in the
past, and about the circumstances leading to its prominence since the late
seventies.  The present article is an outgrowth of this idle curiosity.  I have
been greatly helped in its preparation by the book on soliton physics written
by Russian colleagues$^1$, the essay on Skyrme by Dalitz$^2$ and
a speech by Skyrme reconstructed by Aitchison$^3$, and have relied on these
sources for information when necessary.

Our recent changed perceptions about topology is well brought out by the
following incident.  Some time in the early part of this year, I received a
book from Physics Today entitled {\it Knots and Physics}.  It is written by the
mathematician Louis Kauffman and Physics Today wanted me to review it.

Now, twenty five years ago, it would have been remarkable to send a book on
knots to a physicist for review.  Indeed, a book with a title {\it Knots and
Physics} would have been considered bizarre by physicists and mathematicians
alike.  Twenty five years takes us back to 1968.  It was a time when the
phenomenon of spontaneous symmetry breakdown was only beginning to be widely
appreciated, and electroweak theory had not yet been fully articulated.
Particle physicists were immersed in studies of symmetry principles, and Ken
Wilson was yet to launch lattice QCD.  Physicists and mathematicians had
cordial, but generally distant relations.  Ideas on topology were far from our
minds.  A knot to me meant no more at that time than what is tied at Hindu
weddings.  True, there were a handful of physicists, like David Finkelstein and
Tony Skyrme, who talked of solitons and fundamental groups.  But they were the
oddities.  We were content with Feynman diagrams and current commutators.

But this was not always so.  The first idea on solitons had already occurred
to Scott Russell in 1842.  He was observing the motion of a boat in a narrow
channel and discovered that the water formed by its wake formed a remarkably
stable structure.  He coined the phrase ``solitary elevation'' while discussing
the phenomenon he witnessed.  I reproduce his report on what he saw below.
\newpage
\centerline{{\it Report on Waves.  By} J. Scott Russell, {\it Esq., M.A.,
 F.R.S. Edin.,}}
\centerline{{\it made to the Meetings in} 1842 {\it and} 1843.}
\medskip
\begin{center}
{\it Members of the Committee}
$\left\{
\begin{array}{l}
  \mbox{Sir John Robison*, {\it Sec. R.S. Edin.}}\\
  \mbox{J. Scott Russell, {\it F.R.S. Edin.}}
\end{array}
\right.$
\end{center}

\begin{quotation}
I believe I shall best introduce this phenomenon by describing the
circumstances of my own first acquaintance with it.  I was observing the motion
of a boat which was rapidly drawn along a narrow channel by a pair of horses,
when the boat suddenly stopped--not so the mass of water in the channel which
it had put in motion; it accumulated round the prow of the vessel in a state of
violent agitation, then suddenly leaving it behind, rolled forward with great
velocity, assuming the form of a large solitary elevation, a rounded, smooth
and well-defined heap of water which continued its course along the channel
apparently without change of form or diminution of speed.  I followed it on
horseback, and overtook it still rolling on at a rate of some eight or nine
miles an hour, preserving its original figure some thirty feet long and a foot
to a foot and a half in height.  Its height gradually diminished, and after a
chase of one or two miles I lost it in the windings of the channel.  Such, in
the month of August 1834, was my first chance interview with that singular and
beautiful phenomenon which I have called the Wave of Translation, a name which
it now very generally bears; which I have since found to be an important
element in almost every case of fluid resistance, and ascertained to be the
type of that great moving elevation of the sea, which, with the regularity of a
planet, ascends our rivers and rolls along our shores.
\end{quotation}

It was not Scott Russell alone who came across topological notions in the last
century.  Sir William Thomson had published his ideas about atoms being
vortices in a fluid in 1867$^1$.  Thomson later became Lord Kelvin.
Kelvin did not like the rigid point-like atoms of chemists.  He very much
wanted to describe them visually as extended structures.  He was much impressed
by Helmholtz's discovery of ``vorticity'' in fluids.  On the basis of
experiments with Tait on smoke rings and analytical results on vortices, he
developed his vortex atom which maintained that atoms are vortices in a perfect
fluid.  Some of the vortex atoms of Kelvin are shown in Fig. 1.  He already had
a good intuition about certain knot invariants and seemed upset that he knew
Riemann's ``Lehrs\"{a}tze aus der Analysis Situs'' only through Helmholtz.

\begin{figure}[hbt]
\begin{center}
\mbox{\psboxto(\hsize;0cm){balfig1.eps}}
\end{center}
{\bf Fig. 1 .} {Kelvin proposed that atoms are vortices in a perfect fluid. The
figure shows some of his vortex atoms.}
\end{figure}

But Kelvin's ideas did not catch on.  Fantasies of his sort soon came to be
regarded as reactionary desires to preserve a dissolving mechanistic world.
The ancien r\'{e}gime was losing its ability to rule in the face of the
revolutionary onslaughts of the emerging relativists and quantum theorists.
Physicists generally soon ceased to be seriously bothered by knots and topology
for many years.

But not all physicists.  The young Dirac had published his remarkable text book
on quantum mechanics in 1930.  Soon thereafter$^4$, he came to recognize
certain basic features of wave functions with implications which sharply
differentiate classical from quantum physics.  I can outline his insights from
a modern perspective as follows$^5$.

The dynamics of a system in classical mechanics can be described by equations
of motion on a configuration space $Q$.  These equations are generally of
second order in time.  Thus if the position $q(t_0)$ of the system in $Q$ and
its velocity $\dot{q}(t_0)$ are known at some time $t_0$, then the equations of
motion uniquely determine the trajectory $q(t)$ for all time $t$.

When the classical system is quantized, the state of a system at time $t_0$ is
not specified by a position in $Q$ and a velocity.  Rather, it is described by
a wave function $\psi$ which in elementary quantum mechanics is a (normalized)
function on $Q$.  The correspondence between quantum states and wave
functions however is not one to one since two wave functions which differ by a
phase describe the same state.  The quantum state of a system is thus an
equivalence class
$\{e^{i\a}\psi|\a~{\rm real}\}$
of normalized wave functions.
The physical reason for this circumstance is that experimental observables
correspond to functions like $\psi^*\psi$ which are insensitive to this phase.

In discussing the transformation properties of wave functions, it is often
convenient to enlarge the domain of definition of wave functions in elementary
quantum mechanics in such a way as to naturally describe all the wave functions
of an equivalence class.  Thus instead of considering wave functions as
functions on $Q$, we can regard them as functions on a larger space
$\hat{Q}=Q \times S^1\equiv\{(q,e^{i\a})\}$.  The space $\hat{Q}$ is obtained
by
associating circles $S^1$ to each point of $Q$ and is said to be a $U(1)$
bundle over $Q$.  Wave functions on $\hat{Q}$ are not completely general
functions on $\hat{Q}$, rather they are functions with the property
$\psi(q,e^{i(\a+\theta)})=\psi(q,e^{i\a})e^{i\theta}$.  [Here we can also
replace
$e^{i\theta}$ by $e^{in\theta}$ where $n$ is a fixed integer.]  The behaviour
of
 wave
functions under the action
$(q,e^{i\a})\rightarrow(q,e^{i\a}e^{i\theta})$ of $U(1)$ is
thus fixed.  Because of this property, experimental observables like
$\psi^*\psi$ are independent of the extra phase and are functions on $Q$ as
they should be.  The standard elementary treatment which deals with functions
on $Q$ is recovered by restricting the wave functions to a surface
$\{q,e^{i\a_0}|q\in \hat{Q}\}$ in $\hat{Q}$ where $\a_0$ has a fixed value.
Suc
h a choice
$\a_0$ of $\a$ corresponds to a phase convention in the elementary approach.

When the topology of $Q$ is nontrivial, it is often possible to associate
circles $S^1$ to each point of $Q$ so that the resultant space
$\hat{Q}=\{\hat{q}\}$ is not $Q \times S^1$ although there is still an action
of $U(1)$ on $\hat{Q}$.  We shall indicate this action by $\hat{q} \rightarrow
\hat{q}e^{i\theta}$.  It is the analogue of the transformation $(q,e^{i\a})
\rightarrow (q,e^{i\a}e^{i\theta})$ we considered earlier.  We shall require
this
action to be free, which means that $\hat{q}e^{i\theta}=\hat{q}$ if and only if
$e^{i\theta}$ is the identity of $U(1)$.  When $\hat{Q}\neq Q \times S^1$, the
$U(1)$ bundle $\hat{Q}$ over $Q$ is said to be twisted.  It is possible to
contemplate wave functions which are functions on $\hat{Q}$ even when this
bundle is twisted provided they satisfy the constraint
$\j(\hat{q}e^{i\theta})=\j(\hat{q})e^{in\theta}$ for some fixed integer $n$.
If
 this
constraint is satisfied, experimental observables, being invariant under the
$U(1)$ action, are functions on $Q$ as we require.  However, when the bundle is
twisted, it does not admit globally valid coordinates of the form $(q,e^{i\a})$
so that it is not possible (modulo certain technical qualifications) to make a
global phase choice, as we did earlier.  In other words, it is not possible to
regard wave functions as functions on $Q$ when $\hat{Q}$ is twisted.  [We are
assuming for ease of presentation here that wave functions are always smooth
functions.  That is not of course always the case.  The significance of
$\hat{Q}$ is that smooth functions on $\hat{Q}$ can provide us with physically
acceptable domains for observables.  Discussions involving domains of operators
tend to be technical.  We will not therefore pursue this remark further here.]

It was a great merit of Dirac that already in his seminal paper of 1931 on the
role of phases in quantum theory$^4$, he isolated a physical system
where $\hat{Q}$ was twisted and these phases had an important role.  This was
the system of a particle with electric charge $e$ and a particle with magnetic
charge $g$.  Now we all know that if there is a magnetic monopole (or monopole
for short) at the origin, the
Maxwell equation $\vec{\nabla}\cdot \vec{B}=4\pi g\d^3(x)$ excludes the
existence of a smooth vector potential for the magnetic induction $\vec{B}$
and causes endless trouble in
formulating quantum theory.  The idea of Dirac was to replace the monopole by a
semi-infinite, infinitely thin current loop [see Fig. 2].  This semi-infinite
loop is often
\begin{figure}[hbt]
\begin{center}
\mbox{\psboxto(0cm;8cm){balfig2.eps}}
\end{center}
{\bf Fig. 2 .} {With a magnetic monopole at the origin,$\vec \nabla .
\vec B=4\pi g\,\,\,\delta ^3 (x)$,
$\vec B=$ magnetic induction. Dirac represented magnetic
monopoles by semi-infinite, infinitely thin current loops (Dirac strings). He
showed that the loop can not be observed if electric charge
$e=\frac {\hbar}{2g}n$,
$(n=0,\pm 1,...)$. Thus if a magnetic monopole exists, electric charge is
quantized.}
\end{figure}
called the Dirac string.  The effect of the string away from itself is that of
a monopole, but that is not so exactly on the string.  Dirac argued that the
effect of this string is undetectable if $e=\frac{\hbar}{2g}n$ (in units where
the speed of light $c$ is 1), $n$ being an
integer.  When that is the case, shifting the string amounts
only to changing the phase of the wave function in a way that cannot be
 observed.
Thus, the position of the string, and hence the string
itself, ceases to be an observable when $n$ is an integer.  In that case, then,
Dirac's approach yields a quantum theory for a charge and a monopole [whereas
it gives only a theory of a charge and a semi-infinite current loop when $n$ is
not an integer].  In this way, Dirac derived the quantization of $eg$ in a
charge-monopole theory.

This paper of Dirac is fundamental to quantum theory, and much of modern
mathematics as well.  On the physical side, it predicts charge quantization in
units of $\frac{\hbar}{2g}$ and charge quantization is a basic experimental
fact.  Further, the charge-monopole system has the remarkable feature that it
can have half-odd integral angular momenta even if the charge and monopole have
integral spin.  We are thus led to understand from Dirac's work and later
developments that composites can
have half-odd integral spin even if its constituents have integral spin.
Dirac's paper also suggests the ideas which later were discovered, developed
and
forcefully articulated by Aharonov and Bohm in the context of their celebrated
effect.  As for mathematics, it must be among the first publications on fibre
bundles, the integer $n$ defining what the mathematicians call the Chern class.
In Moli\`{e}re's book, {\it Le Bourgeois Gentilhomme}, M. Jourdain somewhere
exclaims to his philosophy teacher that he was making prose for more than forty
years without knowing it.  The following is a reproduction of the relevant
passage.
\begin{verse}
M. Jourdain: Quoi?  quand je dis: `Nicole, apportez-moi mes pantoufles, et\\
me donnez mon bonnet de nuit', c'est de la prose~?

Ma\^{i}tre de Philosophie: Oui, Monsieur.

M. Jourdain:  Par ma foi!  il y a plus de quarante ans que je dis de la prose\\
sans que j'en susse rien.

M. Jourdain:  What?  when I say: `Nicole, bring me my slippers, and give me \\
my night-cap', is that prose~?

Philosophy Teacher:  Yes, Sir.

M. Jourdain:  Good heavens!  For more than forty years I have been speaking\\
prose without knowing it.

{\it Le Bourgeois Gentilhomme} (1670), II. iv.
\end{verse}

\noindent It would seem that Dirac was unknowingly making mathematics in the
same way that M. Jourdain was unconsciously making prose.

All this happened in 1931.  But we knew nothing about Dirac's work when we
 learned
quantum theory in Madras.  Indeed, quantum theory is still taught in
universities missing out on all the beautiful and fundamental discoveries
coming from the youthful Dirac of 1931.

Dirac's paper, I suppose, must have been too difficult for physicists for
several decades.  What is surely true is that it was largely ignored till the
'70's.  Then events began to unfold in particle and condensed matter theory
bringing topological issues to centerstage, and reviving forgotten memories
of Dirac.  But before describing these events, let us first go back to the late
'50's and the '60's.  At that time, Tony Skyrme and David Finkelstein had also
discovered novel topological ideas which they were developing independently.
They were working alone, or with a colleague or two.  Their ideas too remained
uncomprehended by the community for many years.

Tony Hilton Royle Skyrme was born in England in 1922 in the house of his
maternal grandparents$^2$.  His maternal great-grandfather Edward Robert
knew and admired Kelvin, and was associated with the construction of the Tidal
Predictor under the direction of Kelvin and Tait.  This machine was for
predicting tides worldwide.  The admiration went to the extent of naming his
son Herbert William Thomson Roberts.  This machine was in the house where Tony
was born.  Tony has said in a speech$^3$ that he was greatly impressed
by the ingenuity of its mechanism.

Tony grew up in a world beset with increasing turbulence.  In 1943, after
Cambridge, he joined the British war effort in making the atomic bomb.  It was
only in 1946 that he began fundamental research.  During 1946-61, he was
associated with Cambridge, Birmingham and Harwell and was engaged in wide
ranging investigations in nuclear physics.  It was this work, especially the
work on nuclear matter and the fluid drop model, which eventually culminated in
his beautiful proposition that nucleons are solitons made of pions.

In the speech of Tony I mentioned above, he has described the reasons behind
his extraordinary suggestions.  He knew of Kelvin from the Tidal Predictor, and
was vaguely aware of Kelvin's vortex atoms.  Like Kelvin, he too desired a
model of the nucleon which was visualizable and extended.  He felt that
fermions
can emerge from self-interacting Bose fields just as bosons arise as bound
states of fermions.  Moved by these imprecise and intuitive desires, Tony began
his work on nonlinear field theories, the sine-Gordon equation and the chiral
model and was led to the proposition that nucleons are twisted topological
lumps of pion fields$^6$.  In his papers, Tony also had initiated ideas
on bosonization, vertex operators, and quantum theories on multiply connected
spaces, all years and years ahead of his time, and all topics of central
interest today.

David Finkelstein had a grasp of topology and differential geometry which was
exceptional for physicists in the '60's.  Like Skyrme, he had understood that
solitons can acquire spin 1/2 from the topology of the configuration space.  He
must have realized what little role relativistic quantum field theory played in
the theory of solitons, and struck by the fact that all existing proofs of the
spin-statistics theorem relied on relativistic quantum field theory.  But
solitons can acquire spin half in nonrelativistic models and can not always be
described by relativistic quantum fields.  He and Rubinstein, I suppose, were
led by such thoughts to seek and find an alternative proof of the
spin-statistics theorem.  Their proof was published in 1968$^7$.  It is
this proof which is important for chiral solitons.  There are grounds to expect
that it is the Finkelstein-Rubinstein approach which will be found
significant in
quantum gravity as well.  An absolutely fundamental result, namely the
spin-statistics theorem, is getting topologized.  Even more striking, it is
still not properly understood.

The closing '60's and the '70's herald the dawn of modern times for theorists.
In condensed matter physics, attention began to focus on vortices in
superconductors
and superfluids, and defects in liquid crystals.  Soon a classification of
defects based on its topological properties emerged.  In particle physics, the
bootstrap theory of Chew was leading to unexpected developments.  According to
Chew, there is nuclear democracy, all particles are bound states of each other
and none is more elementary than another.  These ideas evolved into string
theory with its explosive implictions for mathematics and mathematical physics.
Physicists in search of the ultimate solution claim that ``strings are ``TOE'',
``TOE'' meaning ``Theory of Everything.''  If that is so, physicists have found
a scientific substitute for God.  QCD and electroweak theory also began to take
shape in the late '60's and the early '70's.  It soon became plausible that
they
had the ability to account for physics at energies less than about 100 GeV, and
perhaps up to much higher energies as well.  Attention turned to unification of
strong and electroweak theories by the construction of ``grand unified''
models.

It was in this ambience that topology was discovered in particle theory.  There
was first the striking discovery that grand unified theories predicted
monopoles and vortices.  Realization soon dawned that the correct approach to
their study involved the ideas put forth by Dirac in 1931 and the defect
theories of condensed matter.  There was then the discovery of instantons in
QCD and its implications for vacuum structure, time reversal violation and
neutron electric dipole moment.  The remarkable observation was also made by 't
Hooft$^8$ that electroweak theory predicted baryon and lepton number
violation, although at that time, it was felt that the effect was too small to
observe.

There was another dimension to this story.  These developments involved intense
cross fertilization between fields.  There were basic contributions to physics
by leading mathematicians.  It involved as well greater interaction between
condensed matter and particle theorists at least because of their shared
interest in defect theory.

The result of all this activity was that we were bombarded with papers on
topology, and by sheer exposure, if not by effort, began to be aware of
topological issues.  For some of us, this was a period of excitement at
learning beautiful new ideas.  It was also a period of hope that
nonperturbative effects with topological roots can now be investigated with
greater ease.  Instantons, vortices and monopoles began to play a role in
papers on phenomenology and have continued to do so to this day.  The startling
suggestion was made that monopoles can catalize proton decay at strong
interaction rates with life times of the order of perhaps $10^{-19}-10^{-20}$
seconds.
The discovery of 't Hooft about baryon and lepton number violation was also
given teeth by the realization that their rates were greatly enhanced at high
temperatures.  It is nowadays widely speculated that this effect has a
significant bearing on the observed baryon number asymmetry in the universe.

It was during this time that we discovered Skyrme's work at Syracuse$^9$.
There was, before us, the paper of Pak and Tze$^{10}$ reviewing Skyrme's
research.  That too suggested to us that it was worth our
while to work on Skyrmions.  It was in this way that we came to write our
papers on Skyrmions and tell Witten about these ideas.  Soon there followed
Witten's remarkable papers$^{11}$.  It did not take much time thereafter
for the general acceptance of Skyrme's ideas.

Topological notions have flourished with extraordinary vigour in particle
physics, and what is at times called physical mathematics, for ten or more
years.  Much of the recent impetus comes from string and conformal field
theories and their relation to complex manifolds.

Topology is not likely to go away from classical and quantum physics in
the foreseeable future.  It is now appreciated that it has at least two
important roles in physics.  Firstly, it can suggest the existence of stable
structures like defects, vortices, monopoles and Skyrmions.  Secondly, it has a
profound influence on the nature of quantum states.  This influence comes about
because, as mentioned previously, the phases of wave functions can have
serious consequences in quantum theory.  This second feature is still poorly
understood.  Even its existence is not widely known even though sixty two years
have passed after Dirac's paper.  There is also a nonabelian generalisation of
these phases which we have not discussed in this article.

There is also an entirely new role topological physics has recently begun to
assume, brought about because fundamental mathematical developments are
nowadays significantly influenced by quantum field theory.  The contributions
to
knot theory by Witten, the discovery of certain mirror manifolds in string
theory and the recent developments in Riemann surface theory are all dramatic
examples of this role of physics in mathematics.  This then is one more reason
for us to anticipate the vitality and longevity of topological trends in
physics.

It is time to come to an end.  Although there are certain prominent names
associated
with topological physics and its developments, it is good to be conscious that
physics is a social activity in which many humans, not all well-remembered,
participate, and that whatever we do necessarily partakes of the collective
knowledge and creativity of the physics community.  From this perspective, all
the advances and insights which have emerged from studying topological aspects
of physics are also ultimately the fruits of labor of generations of
physicists.  I conclude this essay by quoting a poem by Brecht which aptly
describes these thoughts.
\bigskip

\centerline{A WORKER READS HISTORY}
\begin{center}
\begin{verse}
\medskip
Who built the seven gates of Thebes~?\\
The books are filled with names of kings.\\
Was it kings who hauled the craggy blocks of stone~?\\
And Babylon, so many times destroyed,\\
Who built the city up each time~?  In which of Lima's houses,\\
That city glittering with gold, lived those who built it~?\\
In the evening when the Chinese wall was finished\\
Where did the masons go~?  Imperial Rome\\
Is full of arcs of triumph.  Who reared them up~?  Over whom\\
Did the Caesars triumph~?  Byzantium lives in song,\\
Were all her dwellings palaces~?  And even in Atlantis of the legend\\
The night the sea rushed in,\\
The drowning men still bellowed for their slaves.

\medskip
Young Alexander conquered India.\\
He alone~?\\
Caesar beat the Gauls.\\
Was there not even a cook in his army~?\\
Philip of Spain wept as his fleet\\
Was sunk and destroyed.  Were there no other tears~?\\
Frederick the Great triumphed in the Seven Years War.  Who\\
Triumphed with him~?

\medskip
Each page a victory, At whose expense the victory ball~?\\
Every ten years a great man,\\
Who paid the piper~?

\medskip
So many particulars.\\
So many questions.
\end{verse}
\end{center}

I am most grateful to Paulo Teotonio for drawing the figures and typing their
legends, and to Carl Rosenzweig and Paulo Teotonio for help with the text.
This work was supported by the Department
of Energy under contract number DE-FG02-85ER40231.


\newpage
\centerline{\large\bf{References}}

A long list of references is not appropriate in an informal
essay of this sort.  Only a limited number of publications are therefore cited.

\begin{enumerate}
\item V.G. Makhankov, Yu. P. Rybakov and V.I. Sanyuk, ``The Skyrme Model,
Fundamentals, Methods, Applications'' [Springer-Verlag (in press)].

This book has been especially useful for information on Kelvin and on his
vortex model.
\item R.H. Dalitz, {\it Int. J. Mod. Phys. A}{\bf 3} (1988) 2719.
\item T.H.R. Skyrme, {\it Int. J. Mod. Phys. A}{\bf 3} (1988) 2745.  This
speech was reconstructed by I.J.R. Aitchison.

References 2 and 3 have been freely used while discussing Skyrme.
\item P.A.M. Dirac, {\it Proc. Roy. Soc.} (London) {\bf A133} (1931) 60.
\item A.P. Balachandran, G. Marmo, B.S. Skagerstam and A. Stern, ``Classical
Topology and Quantum States'' [World Scientific, 1991].
\item T.H.R. Skyrme, {\it Proc. Roy. Soc.} (London) {\bf A247} (1958) 260;
{\bf A260} (1961) 121; {\bf A262} (1961) 237; {\it Nucl. Phys.} {\bf 31} (1962)
556; {\it J. Math. Phys.} {\bf 12} (1971) 1735.
\item D. Finkelstein and J. Rubinstein, {\it J. Math. Phys.} {\bf 9} (1968)
1762.  For recent developments involving the Finkelstein-Rubinstein ideas, see
R.D. Tscheuschner, {\it Int. J. Theor. Phys.} {\bf 28} (1989) 1269 [Erratum:
{\it Int. J.
Theor. Phys.} {\bf 29} (1990) 1437]; A.P. Balachandran, A. Daughton, Z-C. Gu,
G. Marmo, A.M. Srivastava and R.D.
Sorkin, {\it Mod. Phys. Letters} {\bf A5} (1990) 1575 and {\it Int. J. Mod.
Phys.} {\bf  A} (in press);
\newline A.P. Balachandran, T. Einarsson, T.R. Govindarajan and R.
Ramachandran,
{\it Mod. Phys. Letters} {\bf A6} (1991) 2801; A.P. Balachandran, W.D. McGlinn,
L.
O'Raifeartaigh, S. Sen and R.D. Sorkin, {\it Int. J. Mod. Phys.} {\bf A7}
(1992)
6887; A.P. Balachandran, W.D. McGlinn, L. O'Raifeartaigh, S. Sen, R.D. Sorkin
and A.M. Srivastava, {\it Mod. Phys. Letters} {\bf A7} (1992) 1427.  These
paper
s
develop proofs of spin-statistics theorems which do not use relativity or
quantum field theory.
\item G. 't Hooft, {\it Phys. Rev. Letters} {\bf 37} (1976) 9; {\it Phys. Rev.}
{\bf D14} (1976) 3432 [Erratum: {\bf D18} (1978) 2199].
\item A.P. Balachandran, V.P. Nair, S.G. Rajeev and A. Stern, {\it Phys.
Rev. Letters} {\bf 49} (1982) 1124; {\it Phys. Rev.} {\bf D27} (1983) 1153.
\item N.K. Pak and H. Ch. Tze, Ann. Phys. (N.Y.) {\bf 117} (1979) 164.
\item E. Witten, {\it Nucl. Phys.} {\bf B223} (1983) 422, 433.

\end{enumerate}
\end{document}